\documentclass[aip,graphicx,floatfix,pop,amsfonts,amssymb,amsmath]{revtex4-1}

\usepackage[utf8x]{inputenc}
\usepackage{hyperref}
\usepackage{todonotes}
\usepackage{upgreek}
\usepackage{fixmath}
\usepackage{lmodern}
\usepackage{subfig}

\renewcommand{\epsilon}{\varepsilon}
\renewcommand{\vec}[1]{\ensuremath{\boldsymbol{#1}}} 

\makeatletter
 \renewcommand{\p@subfigure}{}
\makeatother
\makeatletter
 \providecommand\phantomcaption{\caption@refstepcounter\@captype}
\makeatother
\newcommand{\figwidth}{3.370in}

\newcommand{\grad}{\nabla}

\newcommand{\unit}[1]{\ensuremath{\,\mathrm{#1}}}
\newcommand{\D}[2]{\ensuremath{\frac{\partial#2}{\partial#1}}}

\newcommand{\Eqref}[1]{Eq.~\eqref{#1}}

\newcommand{\Figref}[1]{Fig.~\ref{#1}}

\newcommand{\Tabref}[1]{Tab.~\ref{#1}}

\newcommand{\mynote}[1]{\ensuremath{\mathrm{^#1}}}

\begin{document}


\title{Gyrokinetic modelling of stationary electron and impurity profiles in tokamaks}
\author{A. Skyman}
\email[]{andreas.skyman@chalmers.se}
\author{H. Nordman}
\email[]{hans.nordman@chalmers.se}
\author{D. Tegnered}
\email[]{tegnered@chalmers.se}
\affiliation{Euratom--VR Association, Department of Earth and Space Sciences, \\ Chalmers University of Technology, SE-412 96 Göteborg, Sweden}
\homepage[]{http://www.chalmers.se/rss/}

\date{\today}

\begin{abstract}
Particle transport due to Ion Temperature Gradient/Trapped Electron (ITG/TE)
mode turbulence is investigated using the gyrokinetic code GENE.
Both a reduced quasilinear (QL) treatment and nonlinear (NL) simulations are
performed for typical tokamak parameters corresponding to ITG dominated
turbulence.

A selfconsistent treatment is used, where the stationary local profiles are
calculated corresponding to zero particle flux simultaneously for electrons and
trace impurities.
The scaling of the stationary profiles with magnetic shear, safety factor,
electron-to-ion temperature ratio, collisionality, toroidal sheared rotation,
triangularity, and elongation is investigated.
In addition, the effect of different main ion mass on the zero flux
condition is discussed.

The electron density gradient can significantly affect the stationary impurity
profile scaling.
It is therefore expected, that a selfconsistent treatment will yield results
more comparable to experimental results for parameter scans where the
stationary background density profile is sensitive.
This is shown to be the case in scans over magnetic shear, collisionality,
elongation, and temperature ratio, for which the simultaneous zero flux
electron and impurity profiles are calculated.

A slight asymmetry between hydrogen, deuterium and tritium with respect to
profile peaking is obtained, in particular for scans in collisionality and
temperature ratio.
\end{abstract}

\maketitle


\section{Introduction}
\label{sec:intro}
It is well known that the shape of the main ion density and impurity profiles
are crucial for the performance of a fusion device.
Inward peaking of the main ion (electron) density profile is beneficial for the
fusion performance since it enhances the fusion power production.
For impurities on the other hand, a flat or hollow profile is preferred, since
impurity accumulation in the core leads to fuel dilution and radiation losses
which degrades performance.

The particle profiles are determined by a balance between particle sources and
particle fluxes, a subject which historically has been given much less
attention than energy transport and the associated temperature profiles.
Hence, electron density profiles are often treated as a parameter in
theoretical studies of transport rather than being selfconsistently calculated.

Turbulent transport in the core of tokamaks is expected to be driven mainly by
Ion Temperature Gradient (ITG) and Trapped Electron (TE) modes.
Impurity transport driven by ITG/TE mode turbulence has been investigated in a
number of theoretical studies.\cite{Frojdh1992, Basu2003, Angioni2005,
    Estrada-Mila2005, Naulin2005, Angioni2006, Guirlet2006, Bourdelle2007,
    Dubuit2007, Nordman2007a, Angioni2008, Nordman2008, Angioni2009a,
    Angioni2009c, Camenen2009, Moradi2009, Fable2010, Fulop2010, Futatani2010,
    Hein2010, Angioni2011, Nordman2011, Skyman2011a, Skyman2011b, Casson2013,
    Skyman2014}
Most work in this area has been focused on either scalings of stationary
electron profiles or on impurity transport using prescribed electron density
profiles.

It is well established theoretically that turbulent particle transport in
tokamaks has contributions from both diagonal (diffusive) and non-diagonal
(convective) terms.
The non-diagonal transport contributions may give rise to an inward pinch which
can support an inwardly peaked profile even in the absence of particle sources
in the core.
The stationary peaked profile is then obtained from a balance between diffusion
and convection.

It is known that the electron density gradient can significantly affect the
stationary impurity profile scaling.\cite{Skyman2011b}
In the present work, therefore, the background electron density and impurity
peaking is treated selfconsistently, by simultaneously calculating the local
profiles corresponding to zero turbulent particle flux of both electrons and
impurities.

Linear and nonlinear gyrokinetic simulations using the code
GENE\footnote{\url{http://www.ipp.mpg.de/~fsj/gene/}} are
employed.\cite{Jenko2000, Merz2008a}
The scaling of the stationary profiles with key plasma parameters like magnetic
shear, temperature ratio and temperature gradient, toroidal sheared rotation,
safety factor, and collisionality is investigated for a deuterium (D) plasma.
The isotope scaling of stationary profiles, for hydrogen (H) and tritium (T)
plasmas, is also studied.

The parameters are taken from the Cyclone Base Case (CBC\cite{Dimits2000}),
but with deuterium as main ions; see \Tabref{tab:parameters} for the main
parameters.
It is an ITG~mode dominated scenario and, though set far from marginal
stability, is an interesting case for study, and is widely used as a testing
ground and benchmark for theoretical and numerical studies.

The rest of the paper is organised as follows: In section~\ref{sec:bg} the
theoretical background is given, including considerations regarding analysis
and numerics; the main results are presented in section~\ref{sec:main}, where
scalings of the stationary profiles for electrons and impurities are presented;
results for background peaking for different main ion isotopes is presented and
discussed in section~\ref{sec:isofx}; finally, in
section~\ref{sec:conclusions}, follow the concluding remarks.

\begin{table}
    \centering
    \caption{\small Parameters for the Cyclone Base Case (CBC).
             \mynote{\dag} denotes derived parameters}
    \label{tab:parameters}
    \begin{tabular}{l| r}\hline\hline
        $r/R$                   & $0.18$ \\
        $\hat{s}$               & $0.796$ \\
        $q_0$                   & $1.4$ \\
        $R/L_{n_{i, e}}$        & $2.22$ \\
        $R/L_{T_{i, e}}$        & $6.96$ \\
        $T_i/T_e$               & $1.0$ \\
        $T_e$                   & $2.85 \unit{k eV}$ \\
        $n_e$                   & $3.51 \cdot 10^{19} \unit{m^{-3}}$ \\
        $B_0$                   & $3.1 \unit{T}$ \\
        $R$                     & $1.65 \unit{m}$ \\
        $\beta$                 & $0$ \\
        $\nu_{ei}$\mynote{\dag} & $0.05$ $c_s/R$ \\
        \hline\hline
    \end{tabular}
\end{table}

\section{Background}
\label{sec:bg}
The local particle transport for species $j$ can be formally divided into its
diagonal and off-diagonal parts
\begin{equation}
    \label{eq:transport}
    \frac{R\Gamma_j}{n_j} =
    D_j\frac{R}{L_{n_j}} + D_{T_j}\frac{R}{L_{T_j}} + R V_{p,j}.
\end{equation}
Here, the first term on the right hand side is the diffusion and the second and
third constitute the off-diagonal pinch.
The first of the pinch terms is the particle transport due to the temperature
gradient (thermo-diffusion) and the second is the convective velocity, which
includes contributions from curvature, parallel compression and roto-diffusion.
In equation~\eqref{eq:transport}, $R/L_{X_j}\equiv -R\grad X_j/X_j$ are the
local gradient scale lengths of density and temperature, normalised to the
major radius ($R$).
In general, the transport coefficients dependent on the gradients, though in
the trace impurity limit the transport is linear in both $R/L_{n_Z}$ and
$R/L_{T_Z}$.
A review of the off-diagonal contributions is given in Ref.~\onlinecite{Angioni2012}.

At steady state, the contributions from the different terms in the particle
transport will tend to cancel, resulting in zero particle flux.
Solving equation~\eqref{eq:transport} for zero particle flux, with
$V_j=D_{T_j}1/L_{T_j} + V_{p,j}$ yields
\begin{equation}
 \label{eq:PF}
 PF_j \equiv \left.\frac{R}{L_{n_j}}\right|_{\Gamma=0} = -\frac{R\,V_j}{D_j},
\end{equation}
which is the steady state gradient of zero particle flux for species $j$.
This measure quantifies the balance between diffusion and advection, and gives
a measure of how ``peaked'' the local density profile is at steady state.
It is therefore referred to as the ``peaking factor'' and denoted $PF_j$.

\begin{figure}[tb]
    \centering
    \subfloat[
        electron particle flux for different density gradients
        \label{fig:flux_times}]{
        \includegraphics[width=\figwidth]{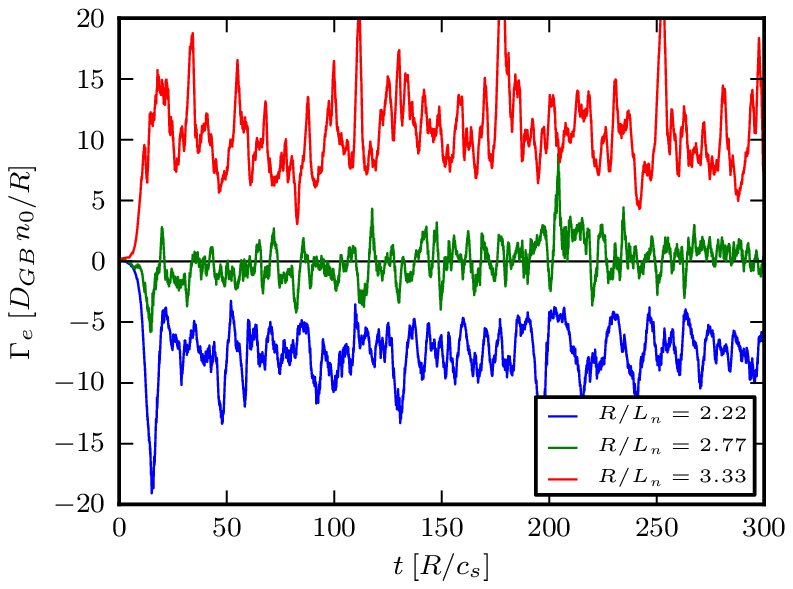}}

    \subfloat[
        poloidal wavenumber spectra of the electron particle flux, evaluated at the gradient of zero flux
        \label{fig:flux_spectrum}]{
        \includegraphics[width=\figwidth]{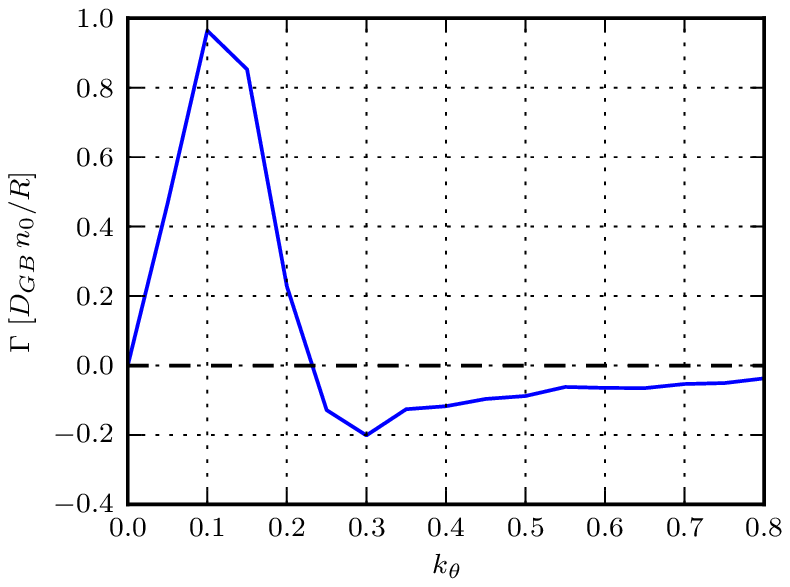}}
    \caption{\small Timeseries and spectra of electron particle transport for CBC with D as main ions
        near to the zero flux gradient ($R/L_{n_e}=2.77$); obtained from NL GENE simulations.}
\end{figure}

In order to investigate the transport, nonlinear (NL) GENE simulations were
performed from which $PF_e$ for the stationary electron profiles were
calculated.
The results were compared with quasilinear (QL) results, also obtained using
GENE.
The background peaking factor was found by explicitly seeking the gradient of
zero particle flux by calculating the electron flux for several values of the
density gradient.
A typical set of simulations is displayed in \Figref{fig:flux_times}, where the
time evolution of the electron flux for three density gradients near the
gradient of zero particle flux is shown (fluxes are in gyro-Bohm units, with
$D_\text{GB}=c_\text{s}\rho_\text{s}^2/R$).
A second order polynomial $p$ was then fitted to the data closest to the zero
flux gradient and then the $PF_e$ was found as the appropriate root of $p$.
The error for $PF_e$ was approximated by finding the corresponding roots of $p
\pm \text{max}\left[\sigma_{\Gamma}\right]$, and using the difference between
these roots as a measure of the error.
In \Figref{fig:flux_spectrum} the particle flux spectrum for a NL simulation
for CBC near this gradient is shown.
The figure illustrates that the total flux is zero due to a balance of inward
and outward transport occurring at different wavenumbers.
The method for finding $PF_e$ from the QL simulations is the same, but here a
reduced treatment was used, including only the dominant mode, which is an ITG
mode for CBC-like parameters.
This was done for a range of values of several key plasma parameters.

In the trace impurity limit, i.e. when the fraction of impurities is
sufficiently small,
the impurity dynamics do not affect the turbulence dynamics.
Therefore, when finding the simultaneous peaking factor of the background and
impurities, the former can be found first and used in the simulations of the
latter without loss of generality.
Furthermore, in the trace impurity limit, the transport coefficients of
\Eqref{eq:transport} for trace impurities do not depend on the species'
gradients of density and temperature, meaning that \eqref{eq:transport} is a linear
relation in those gradients.
This means that the impurity peaking, as well as the contribution to $PF_Z$ from
the thermodiffusion ($PF_T$) and the convective velocity ($PF_p$), can be found
from simulations with appropriately chosen gradients using the method outlined
in~\cite{Casson2010}.
The peaking factors are calculated for several impurity species, using the
reduced QL model.
The difference in impurity peaking factors between NL and QL models has been
covered in previous work.\cite{Nordman2011, Skyman2011a, Skyman2011b,
Skyman2014}

The simulations have been performed in a circular equilibrium with aspect ratio
$R/a=3$, using kinetic ions, electrons and impurities, except when studying the
effects of shaping.
Then the Miller equilibrium model was used instead.\cite{Miller1998}
Impurities were included at trace amounts ($n_Z/n_e=10^{-6}$), so as not to
affect the turbulent dynamics.
The impurities mass was assumed to be $A_Z=2Z$, where $Z$ is the charge
number.
The dynamics were further assumed to be electrostatic ($\beta \approx 0$).

For the simulation domain, a flux tube with periodic boundary conditions in the
perpendicular plane was used.
The nonlinear simulations were performed using a $96\times 96\times 32$ grid in
the normal, bi-normal, and parallel spatial directions respectively; in the
parallel and perpendicular momentum directions, a $48\times 12$ grid was used.
For the linear and quasilinear computations, a typical resolution was $12\times
24$ grid points in the parallel and normal directions, with $64\times
12$ grid points in momentum space.
The nonlinear simulations were typically run up to $t=300\, R/c_\text{s}$ for
the experimental geometry scenario, where $R$ is the major radius and
$c_\text{s}=\sqrt{T_e/m_i}$.

\section{Simultaneous stationary profiles of electrons and impurities}
\label{sec:main}
\begin{figure}[tb]
    \centering
    \includegraphics[width=\figwidth]{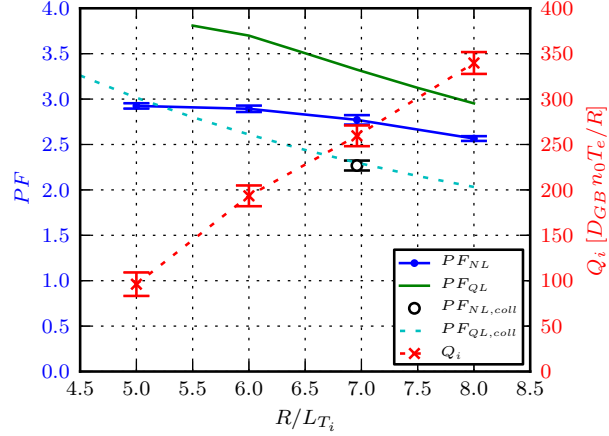}
    \caption{\small
    Scaling of $PF_e$ and ion heat flux with ion temperature gradient $\grad T_i$.}
        \label{fig:NL_QL_omt}
\end{figure}

First, we examine the dependence of the transport and of $PF_e$ on the ion
temperature gradient.
The result is shown in \Figref{fig:NL_QL_omt}, where the ion energy transport
from NL simulations is displayed, together with electron peaking factors from
NL and QL simulations.
Though the ion energy transport shows a stiff increase with the driving
gradient, only a moderate reduction is seen in the peaking factor.

\begin{figure}[tb]
    \subfloat[
        NL and QL scalings of $PF_e$
        \label{fig:Tj_NL_QL_coll}
    ]{\includegraphics[width=\figwidth]{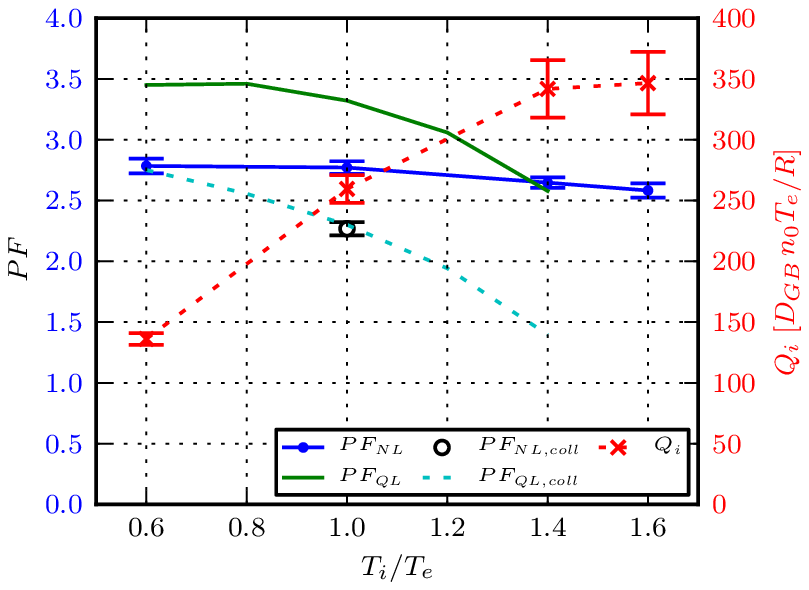}}~
    \subfloat[
        simultaneous QL scalings of $PF_e$ and $PF_Z$
        \label{fig:Tj_imps}
        ]{\includegraphics[width=\figwidth]{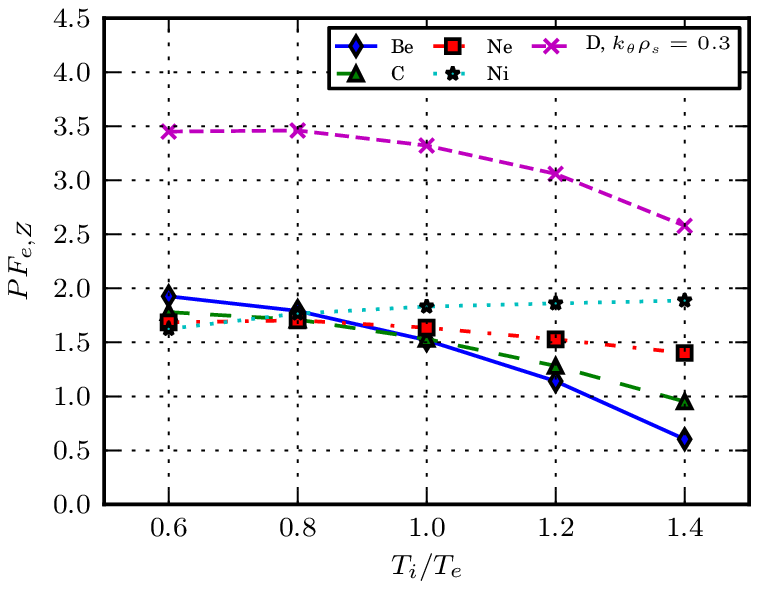}}

    \subfloat[
        contributions to $PF_Z$ from thermopinch ($PF_T$) and pure convection ($PF_p$) vs. impurity charge
        \label{fig:Tj_parts}
        ]{\includegraphics[width=\figwidth]{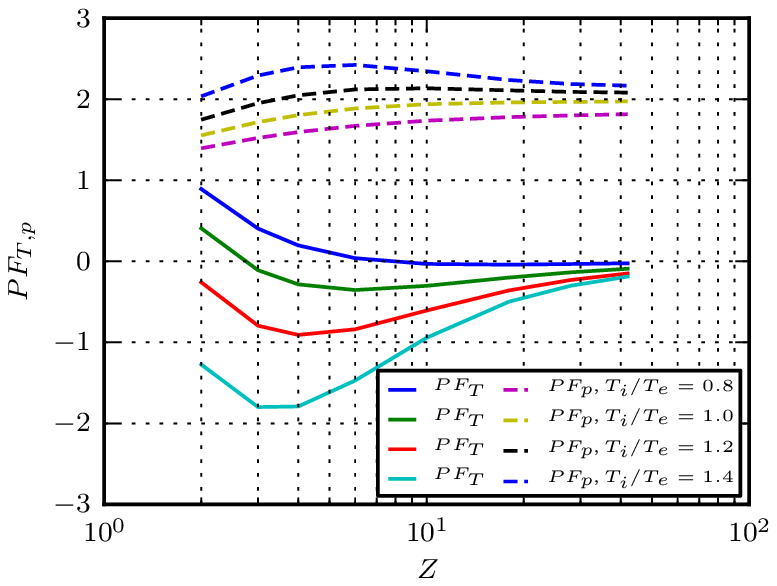}}~
    \subfloat[
        scaling of ITG growthrates ($\gamma$) and real frequencies ($\omega_r$),
        normalised to $c_\text{s}/R$
        \label{fig:Tj_spectra}
        ]{\includegraphics[width=\figwidth]{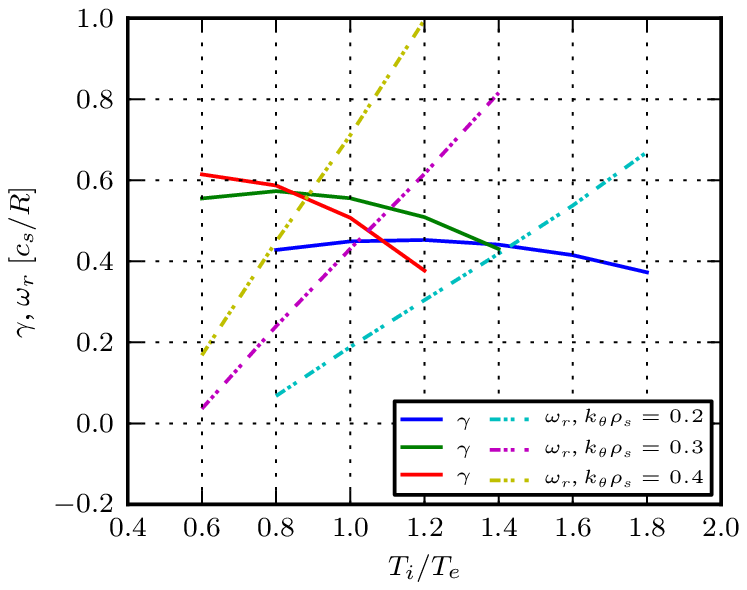}}

    \caption{\small Scaling of background electron peaking, impurity peaking and
            linear eigenvalues with $T_i/T_e$.}
    \label{fig:Tj_NL_QL}
\end{figure}

\begin{figure}[tb]
    \subfloat[
        NL and QL scalings of $PF_e$
        \label{fig:shat_NL_QL_coll}
    ]{\includegraphics[width=\figwidth]{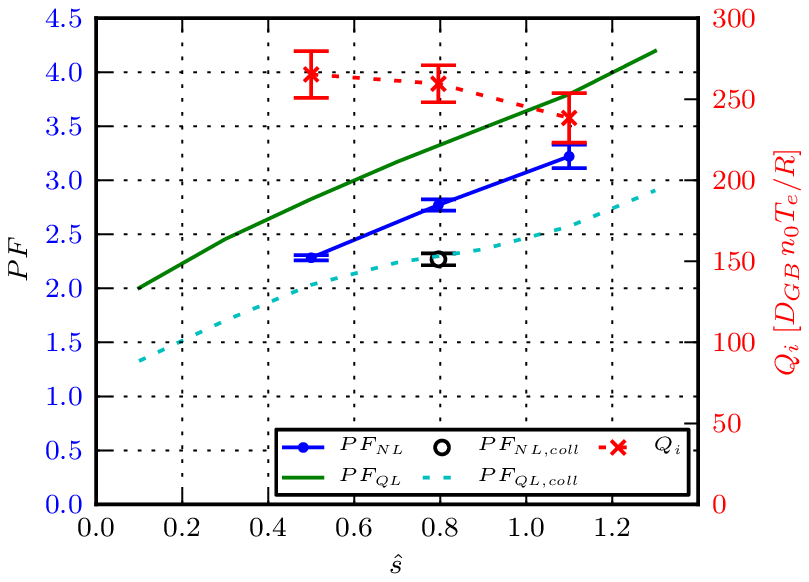}}
    ~
    \subfloat[
        simultaneous QL scalings of $PF_e$ and $PF_Z$
        \label{fig:shat_imps}
        ]{\includegraphics[width=\figwidth]{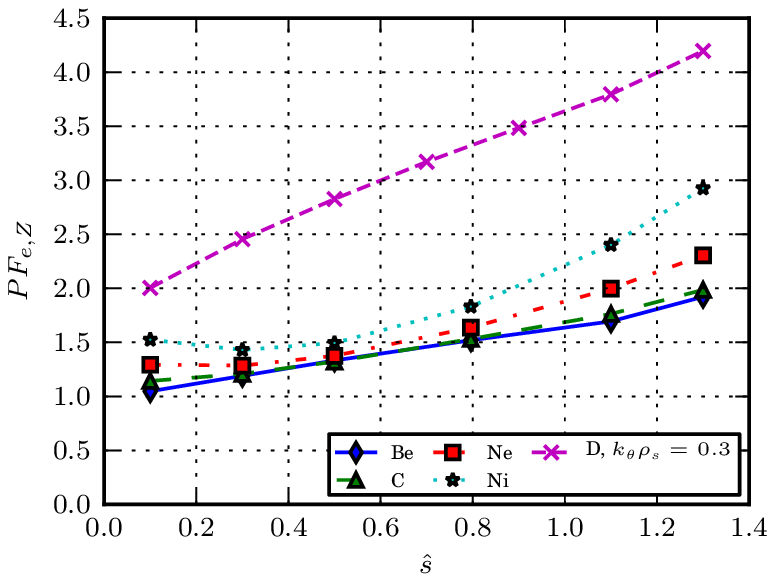}}

    \subfloat[
        contributions to $PF_Z$ from thermopinch ($PF_T$) and pure convection ($PF_p$) vs. impurity charge
        \label{fig:shat_parts}
        ]{\includegraphics[width=\figwidth]{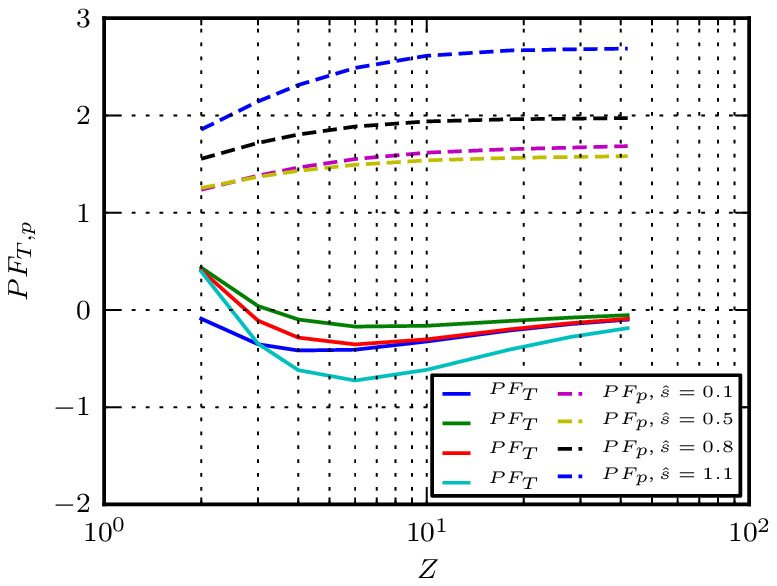}}

    \caption{\small Scaling of background electron  and impurity peaking with
        $\hat{s}$.}
    \label{fig:NL_QL_shat}
\end{figure}

\begin{figure}[tb]
    \centering
    \subfloat[
        simultaneous QL scaling of $PF_e$ and $PF_Z$
        \label{fig:coll_imps}]{
        \includegraphics[width=\figwidth]{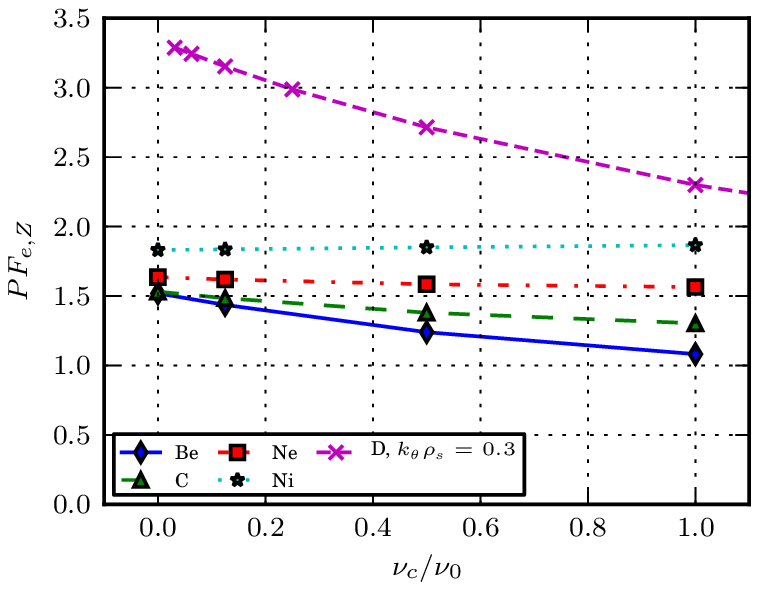}}

    \subfloat[
        contributions to $PF_Z$ from thermopinch ($PF_T$) and pure convection ($PF_p$) vs. impurity charge
        \label{fig:coll_parts}
        ]{\includegraphics[width=\figwidth]{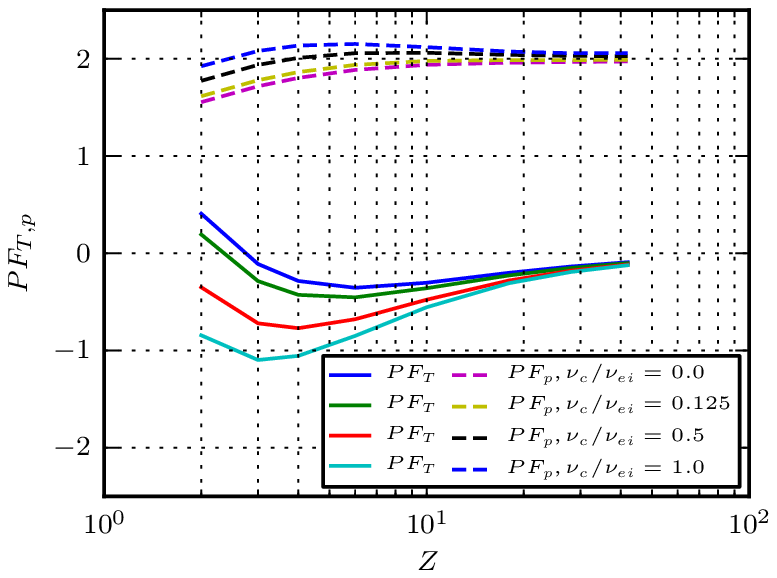}}

    \caption{\small Scaling of background electron  and impurity peaking with
        $\nu_{ei}$.}
    \label{fig:coll}
\end{figure}

\begin{figure}[tb]
    \centering
    \subfloat[
        simultaneous QL scaling of $PF_e$ and $PF_Z$
        \label{fig:Vtor_imps}]{
        \includegraphics[width=\figwidth]{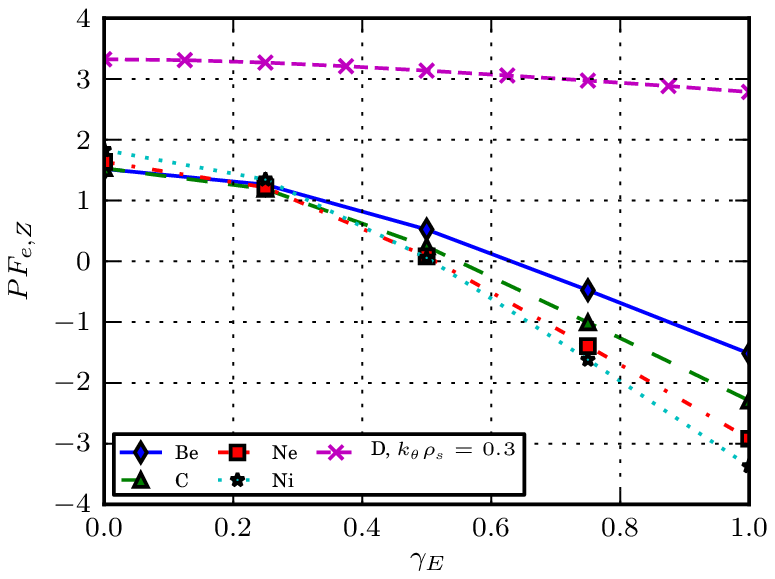}}

    \subfloat[
        contributions to $PF_Z$ from thermopinch ($PF_T$) and pure convection ($PF_p$) vs. impurity charge
        \label{fig:Vtor_parts}
        ]{\includegraphics[width=\figwidth]{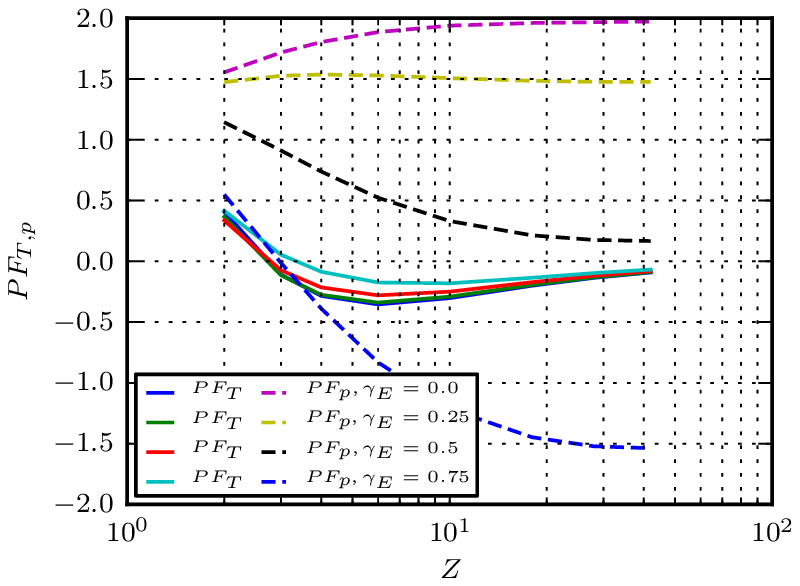}}

    \caption{\small Scaling of background electron  and impurity peaking with
        $\gamma_E$.}
    \label{fig:Vtor}
\end{figure}

\begin{figure}[tb]
    \centering
    \subfloat[
        simultaneous QL scaling of $PF_e$ and $PF_Z$
        \label{fig:kappa_imps}]{
        \includegraphics[width=\figwidth]{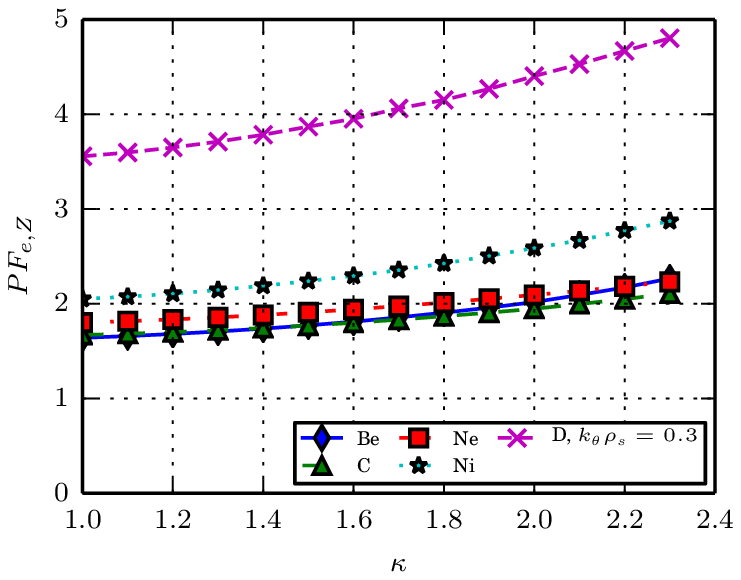}}

    \subfloat[
        contributions to $PF_Z$ from thermopinch ($PF_T$) and pure convection ($PF_p$) vs. impurity charge
        \label{fig:kappa_parts}
        ]{\includegraphics[width=\figwidth]{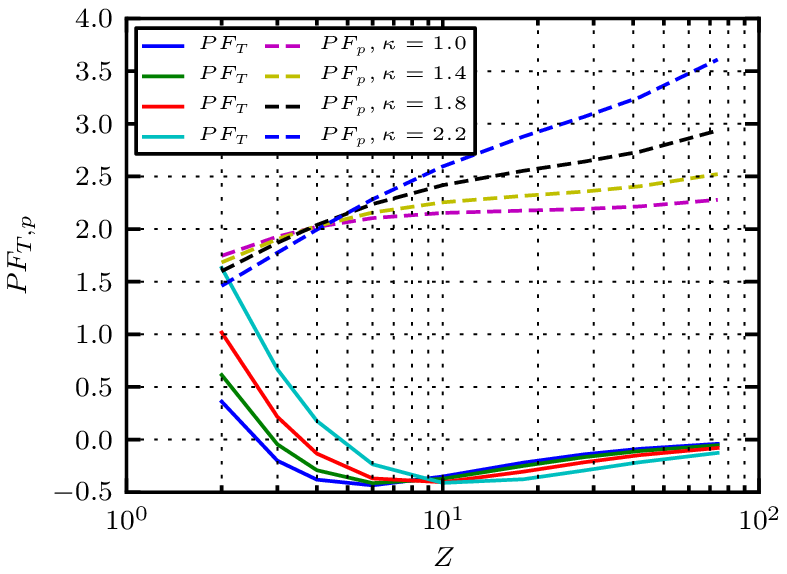}}

    \caption{\small Scaling of background electron and impurity peaking with
        $\kappa$.}
    \label{fig:kappa}
\end{figure}

It is worth noting that the steady state peaking found in the simulations is
considerably higher than that in the original CBC experiment
($R/L_{n_{e,i}}=2.22$).
As is known\cite{Angioni2005, Angioni2009b, Angioni2009c, Fable2010}, this is
due to the neglect of collisions, as they normally are in the CBC.
The collisionality for the CBC parameters is $\nu_{ei} \approx 0.05\,
c_\text{s}/R$, which is of the same order as the growthrates and real
frequencies observed, and collisions can be expected to have a notable impact
on the transport.
When collisions are added, the background peaking factor is indeed lowered to a
level consistent with the prescribed background gradient for the CBC, as seen
in \Figref{fig:NL_QL_omt}.
The QL peaking factor shows a stronger decrease than its NL counterpart.
Below $R/L_{T_i}\approx 4.5$, the ITG~mode is stable, and the TE~mode
dominates.
In the following, focus is on the collisionless case, but the simulations have
been complemented with scalings including collisions.

The electron peaking factor is reduced with increasing ion--electron
temperature ratio ($T_i/T_e$) for CBC parameters, as can be seen in
\Figref{fig:Tj_NL_QL_coll}.
As with the temperature gradient, the NL results show only a weak
scaling, while the trend is more pronounced for the QL simulations.
This may be a result of the QL treatment, which only includes the dominant
mode, while the contribution from the subdominant TE~mode is non-negligible for
low values of $T_i/T_e$.
A more complete QL treatment may give a better
agreement.\cite{Bourdelle2007, Fable2010}
In \Figref{fig:Tj_imps}, the selfconsistently obtained quasilinear peaking of
electrons and impurities (Be ($Z=4$), C ($Z=6$), Ne ($Z=10$), and Ni ($Z=28$))
is shown.
Impurities with lower charge numbers ($Z$), as well as the background, show
the same dependence on $T_i/T_e$, with a decrease in the peaking as the ion
temperature is increased, and a weaker tendency for smaller wavenumbers.
For the impurities with higher $Z$, on the other hand, increased ion
temperature leads to slightly more peaked impurity profiles.
In \Figref{fig:Tj_parts} it is shown that the effect for the impurities is
mainly due to an increase in the relative contribution from the outward
thermopinch ($\sim1/Z$) with increased ion temperature, which affects the low
$Z$ impurities more strongly.
To first order, the thermopinch is proportional to the real frequency.
As seen in~\Figref{fig:Tj_spectra} it increases with increasing $T_i/T_e$,
which explains its increasing importance for higher ion temperatures.

In \Figref{fig:shat_NL_QL_coll}, the scaling with magnetic shear ($\hat{s}$) is
studied.
The electron peaking shows a strong and near linear dependence on $\hat{s}$.
This is similar to the results reported in Ref.~\onlinecite{Fable2010} and is due
to the shear dependence of the curvature pinch.
This trend is as strong in both the QL and NL simulations.
The effect of shear on the linear eigenvalues is not monotonous, with
a destabilisation in the low to medium shear region followed by stabilisation
as $\hat{s}$ is increased further.
The selfconsistent results are shown in \Figref{fig:shat_imps}.
For the impurities, the change in peaking factors due to magnetic shear follows
the trend seen for the electrons, and impurities with higher $Z$ are more
strongly affected.
This is seen in~\Figref{fig:shat_parts} to be due mainly to a stronger inward
convective pinch with increasing shear.



Next we cover the effect of electron--ion collisions on the peaking factors.
Collisionality is known to affect the background by reducing the peaking
factor.\cite{Angioni2005, Angioni2009a, Angioni2009c, Fable2010}
In \Figref{fig:coll_imps}, the selfconsistent results for a range of
collisionalities are shown.
The reduction in peaking factors with collisionality is also seen for low $Z$
impurities, while the high-$Z$ impurities show little or no change in peaking
due to collisions.
The effect on the impurities is mainly due to an increase in the outward
thermopinch ($\sim1/Z$) with increased collisionality
(\Figref{fig:coll_parts}), due to a change of the real frequency.

The influence of sheared toroidal flows on the selfconsistent impurity
peaking was also studied.
Only purely toroidal rotation was considered, included through the
$\vec{E}\times\vec{B}$ shearing rate, defined as $\gamma_E =
-\frac{r}{q}\frac{1}{R}\D{r}{v_\text{tor}}$.
Hence, we flow shear in the limit where the flow is
small, neglecting effects of centrifugal and Coriolis forces.
These may, however, be important for heavier impurities.\cite{Camenen2009}
The results are shown in \Figref{fig:Vtor_imps}, where it can be seen that
impurities are much more strongly affected by the rotation than the electrons,
due to the difference in  thermal velocity.
For large values of $\gamma_E$, a strong decrease in impurity peaking is seen.
The effect is due to the outward roto-pinch which becomes important for large
values of $\gamma_E$, as shown in \Figref{fig:Vtor_parts}.
As with the shearing rate, this effect is more pronounced for high-$Z$
impurities, since the thermopinch dominates for low $Z$ values.
In ASDEX~U roto-diffusion has been found to be a critical ingredient to
include in order to reproduce the Boron profiles seen in
experiments.\cite{Angioni2011, Casson2013}

Finally, shaping effects were studied using the Miller equilibrium model.
The quasi-linear electron peaking factor as well as the self-consistent
impurity peaking factors increase with higher elongation  $(\kappa)$ as shown
in \Figref{fig:kappa_imps}.
For impurities with low charge number the increase in peaking is mainly due to
a larger inward thermopinch while for high-$Z$ impurities it is caused by an
increased pure convection, as seen in \Figref{fig:kappa_parts}.

The dependence of the selfconsistent peaking factors on the safety factor
($q_0$) and triangularity ($\delta$) was also studied, and the scalings were
found to be very weak.



\section{Isotope effects on the background peaking}
\label{sec:isofx}
The CBC prescribes hydrogen ions as the main ions, however, for future fusion
power plants, a deuterium/tritium mixture will be used.
Due to the difference in mass, it is known that D and T plasmas will behave
differently from pure H plasmas.
Differences in steady state peaking factors are expected, since both collisions
and non-adiabatic electrons can break the gyro-Bohm scaling.\cite{Pusztai2011}
To get an insight into the effect of the main ion isotope, the scalings for the
normal CBC were compared with simulations where D was substituted for H and T.

\begin{figure}[tb]
    \centering
    \includegraphics[width=\figwidth]{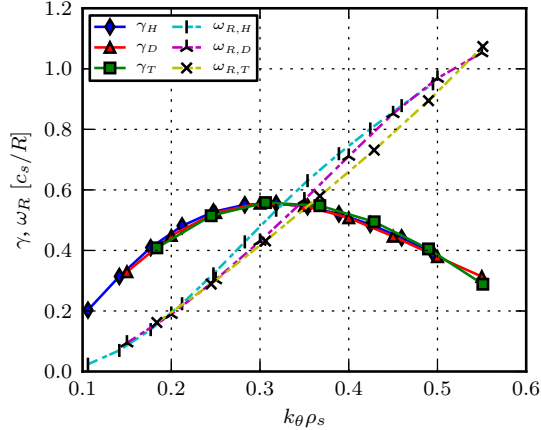}
    \caption{\small
             Eigenvalue spectra for CBC parameters (\Tabref{tab:parameters}), for
             H, D and T as main ions, with $k_\theta\rho_{\text{s}i}$ and eigenvalues
             in species units ($c_{\text{s}i}/R$).}
    \label{fig:spectra}
\end{figure}

\begin{figure}[tb]
    \centering
    \subfloat[%
        scaling with collisionality
        \label{fig:PF_coll_HDT}
    ]{\includegraphics[width=\figwidth]{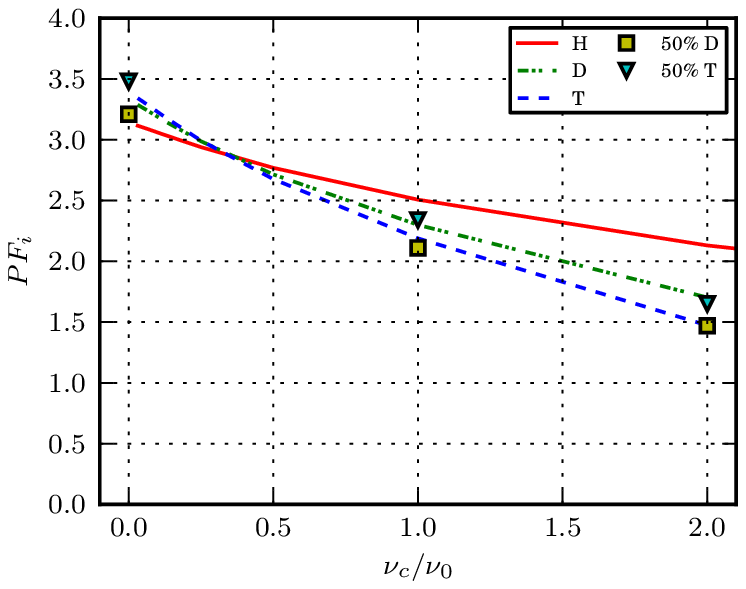}} \\
    \subfloat[%
        scaling with temperature ratio
        \label{fig:PF_TiTe_HDT}
    ]{\includegraphics[width=\figwidth]{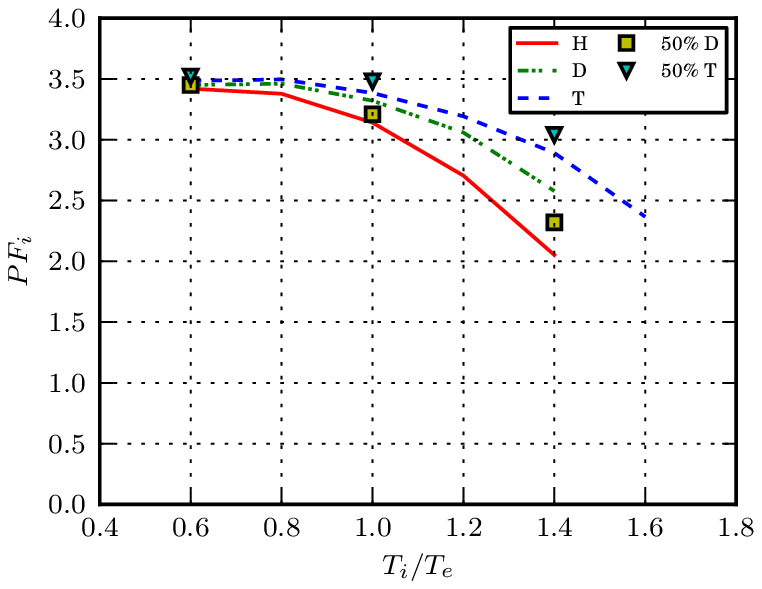}}

    \caption{\small
             Scaling of main ion peaking with different parameters for the CBC
             (\Tabref{tab:parameters}), for H, D and T as main ions with
             $k_\theta\rho_{\text{s}i}=0.3$ in species units.}
    \label{fig:PF_main}
\end{figure}

\begin{figure}[tb]
    \centering
    \includegraphics[width=\figwidth]{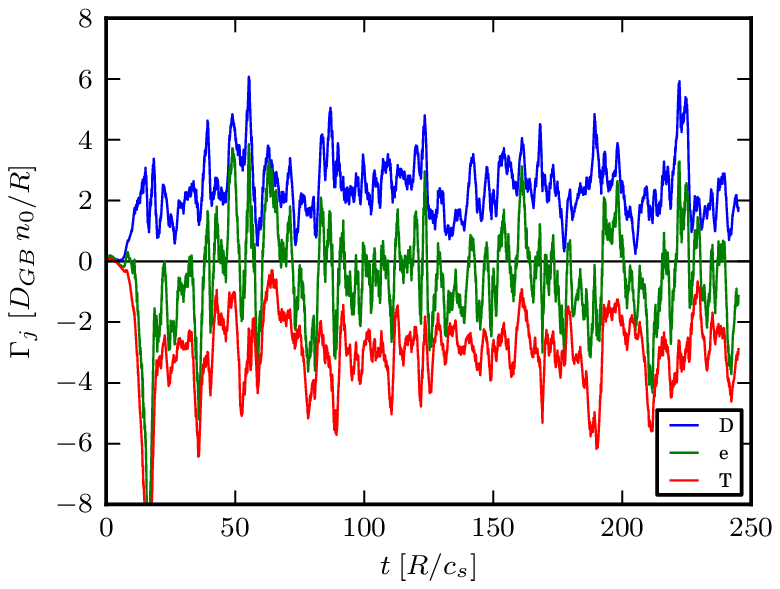}
    \caption{\small
        Timeseries of D, T and $e$ particle flux for CBC
        (\Tabref{tab:parameters}) with a 50/50 mixture of D and T as main ions.
        Evaluated at the zero flux gradient for the pure D
        case ($R/L_{n_e}=2.77$).
    }
    \label{fig:time_DT5050}
\end{figure}

First, we review the known isotope effects on linear eigenvalues.
Figure~\ref{fig:spectra} displays the ITG eigenvalues in the collisionless case
for H, D, and T in species units.
The slight difference in eigenvalues obtained is due to the non-adiabatic
electron response into which the mass ratio $\sqrt{m_i/m_e}$ enters, as
discussed in Ref.~\onlinecite{Pusztai2011}.

The QL background peaking versus collisionality is displayed in
\Figref{fig:PF_coll_HDT} for $k_\theta\rho_{\text{s}i}=0.3$ in species units,
corresponding to the peaks in the growthrate spectra.
For $\nu_{ei}=0$, a slight difference in $PF$ is observed, with $PF_\text{T} >
PF_\text{D} > PF_\text{H}$.
This is consistent with the asymmetry in D and T transport reported in
Refs.~\onlinecite{Estrada-Mila2005, Nordman2005}.
For larger values of the collisionality, however, the order is reversed.

Next, the effect of the ion mass on the stationary profile scaling with ion to
electron temperature ratio ($T_i/T_e$) is studied.
In \Figref{fig:PF_TiTe_HDT}, the peaking factor is seen to decrease with
increasing ion temperature, but in this case the lighter isotopes are more
sensitive, showing a stronger decrease with $T_i/T_e$.
The other parameter scalings discussed in section~\ref{sec:main} show only a
very weak isotope effect.

The scenario with a 50/50 mixture of D and T was also studied, and the
simultaneous peaking of D and T calculated.
The results were seen to follow the pure D and pure T results closely,
albeit with the T profile approximately 10\% more peaked than the D profile for
all values of the collisionality; see \Figref{fig:PF_coll_HDT}.
For the scan with $T_i/T_e$, the self-consistent case gave a larger difference
in D and T peaking than the corresponding pure cases, as seen in
\Figref{fig:PF_TiTe_HDT}.
These results were corroborated by NL simulations using the standard CBC
parameters, with the background electron density gradient corresponding to zero
flux for the pure D case ($R/L_{n_{e}}=2.77$).
The results are shown in \Figref{fig:time_DT5050}.
For these parameters, the electron particle flux remained close to zero, while
the deuterium flux was postivive and the tritium flux negative, indicating a
more peaked steady state D profile, and a less peaked T profile in the mixed
scenario.

The effect of main ion mass on the stationary profiles discussed here are
weak, but may result in a D--T fuel separation in a fusion
plasma.\cite{Estrada-Mila2005}


\section{Conclusions}
\label{sec:conclusions}
In the present paper electron and impurity particle transport due to Ion
Temperature Gradient/Trapped Electron (ITG/TE) mode turbulence was studied
using gyrokinetic simulations.
A reduced quasilinear (QL) treatment was used together with nonlinear (NL)
simulations using the code GENE.
Neoclassical contributions to the impurity transport, which may be relevant for
high-$Z$ impurities, were neglected.
The impurities, with impurity charge in the region $3 \le Z \le 42$, were
included in low concentrations as trace species.
The focus was on a selfconsistent treatment of particle transport, where the
stationary local profiles of electrons and impurities are calculated
simultaneously corresponding to zero particle flux.
The zero flux condition is relevant to the core region of tokamaks where the
particle sources are absent or small.
The parameters were taken from the Cyclone Base Case, corresponding to ITG
dominated turbulence with a subdominant TE mode relevant for the core region of
tokamaks, and scalings of the stationary profiles with magnetic shear, safety
factor, electron-to-ion temperature ratio, collisionality, sheared toroidal
rotation, elongation and triangularity were investigated.

It was shown that the stationary background density profile was sensitive in
scans over magnetic shear, collisionality, elongation, and temperature ratio,
for which the simultaneous zero flux electron and impurity profiles are
calculated.
The selfconsistent treatment mainly tended to enhance these parameter scalings
of the impurity profile peaking.
For safety factor, sheared toroidal rotation and triangularity on the other
hand, the effects on the electron profile were weak and hence a selfconsistent
treatment did not add significant new results to the previous investigations in
this area.
For all considered cases, both the electron profile and the impurity profile
were found to be inwardly peaked, with peaking factors $R/L_{n_Z}$ typically in
the range $1.0$--$4.0$, i.e. substantially below neoclassical expectations.
For large sheared toroidal rotation ($\gamma_E \gtrsim 0.4$), a flux reversal
resulting in outwardly peaked impurity profiles was seen.
Furthermore, the electrons were consistently more peaked than the impurities.

In addition, a  slight asymmetry between hydrogen, deuterium and tritium with
respect to profile peaking was obtained.
The effect was more pronounced for high collisionality plasmas and large ion to
electron temperature ratios.
The effect may have consequences for fuel separation in D--T fusion plasmas.

\section*{Acknowledgements}
\label{sec:acknowledgements}
This work was funded by a grant from The Swedish Research Council (C0338001).

The main simulations were performed on resources provided on the
Lindgren\footnote{See
\protect\url{http://www.pdc.kth.se/resources/computers/lindgren/} for details
on Lindgren} high performance computer, by the Swedish National Infrastructure
for Computing (SNIC) at Paralleldatorcentrum (PDC).

The authors would like to thank F~Jenko, T~Görler, MJ~Püschel, D~Told, and the
rest of the GENE~team at IPP--Garching for their valuable support and input.



%

\end{document}